\documentclass[12pt]{iopart}
\pdfoutput=1
\usepackage{graphicx}
\usepackage{bm}
\usepackage{hyperref}
\usepackage{color}
\usepackage{soul}

\usepackage{cite}

\begin{document}
\title{Optimizing the Throughput of Particulate Streams Subject to Blocking}
\author{G. Page$^1$, J. Resing$^2$, P. Viot$^1$ and J. Talbot$^1$}
\address{
$^1$Laboratoire de Physique Th\'eorique de la Mati\`ere Condens\'ee, CNRS  UMR 7600, Sorbonne Universit{\'e}s,
 4, place Jussieu, 75252 Paris Cedex 05, France.\\
$^2$Department of Mathematics and Computer Science, Eindhoven University of Technology, Netherlands.
}

\date{\today}
\begin{abstract}
Filtration, flow in narrow channels and traffic flow are examples of processes subject to blocking
when the channel conveying the particles becomes too crowded. If the blockage is temporary, which means that after 
a finite time the channel is flushed and reopened, one expects
to observe a maximum throughput for a finite intensity of entering particles. We investigate this phenomenon
by introducing a queueing theory inspired, circular Markov model. 
Particles enter a channel with intensity $\lambda$ and exit at a rate $\mu$. If
$N$ particles are present at the same time in the channel, the system becomes blocked and no more particles can enter until 
the blockage is cleared after an exponentially distributed time with rate $\mu^*$. 
We obtain an exact expression for the steady state throughput (including the exiting blocked particles) for all values of $N$.
For $N=2$ we show that the throughput assumes a maximum value for finite $\lambda$ if
$\mu^*/\mu < 1/4$. The time-dependent throughput either monotonically approaches the steady state value, or reaches a maximum value at finite time. 
We demonstrate that, in the steady state, this model can be mapped to a previously introduced non-Markovian model with fixed transit and blockage times. 

We also examine an irreversible, non-Markovian blockage process with constant transit time exposed to an entering flux
of fixed intensity for a finite time and we show that the first and second moments of the number of exiting particles are maximized for a finite intensity.

{\bf Keywords:} exact results; non-equilibrium processes
\end{abstract}
\maketitle

\section{Introduction}

Whenever the carrying capacity of a channel is limited, blockage is a possibility
and indeed the phenomenon is commonly observed over a range of length scales \cite{Zuriguel2011}. 
Examples include filtration \cite{Redner2000,Roussel2007} vehicular and pedestrian
traffic flow \cite{Helbing2001,Nagatani2002,tajima2001scaling}, 
granular systems, the flow of macro-molecules through micro or nanochannels \cite{Finkelstein1981,Kapon2008,marin2018clogging} and in other applications
like internet attacks (DoS) \cite{bhunia2014}. 
The blockage may be temporary, in which case the flow
resumes after a certain amount of time has elapsed, or it may be irreversible with no reopening of the channel
possible. 

In the first situation, if the entering flux is constant, 
a steady state with alternating open and
blocked states will eventually be reached \cite{BTV2013}. Increasing the intensity of
entering particles would, in the absence of blockage, lead to a proportionate
increase in the throughput, or rate of
exiting particles. With blockage present, however, increasing the intensity
increases the probability of blockage that disrupts the
throughput. Thus, one can expect that, under certain conditions, the throughput will be maximized
for a finite intensity of entering particles. 

When the blockage is irreversible, the total number of exiting
particles can be continually increased by reducing the intensity to values approaching zero \cite{Talbot2015}. 
This will, of course, require an ever-increasing amount of time.
A more interesting question is to consider a situation in which the entering flux ceases at  a
giving stopping time. Here we expect that the total number of exiting
particles can be optimized for a finite intensity.  

Such considerations also arise in queueing theory that has long been used to analyze
service operations performed on units arriving according to a given distribution \cite{huntseq,medhi2002stochastic}. Traditional applications include industrial engineering, telecommunications and traffic flow. More recently, 
queueing theory has been used in biophysics, for example to model enzymatic servers \cite{cookson2011queueing}.
In many of these applications, the throughput of serviced jobs is a crucial quantity. An example taken from industrial engineering is a closed loop conveyor system with homogeneous servers \cite{elsayed1983multichannel}.

Here we introduce a circular Markov chain model with a blocked state that arises when $N$ particles are simultaneously present in the system. The blockage is subject to removal at a constant rate. 
We demonstrate that the throughput can be maximized (i) at a finite intensity of entering particles in the steady state; (ii) as a function of time in the 
transient regime; (iii) in the transient regime only or (iv) in neither, depending on the rates of transit and deblockage. We also compare the model to a previously introduced
semi-deterministic process with constant transit and blockage times as well as with the Erlang loss formula.

\section{Circular Markov Model of Reversible Blockage}

\begin{figure}
\begin{center}
\includegraphics[width=12cm]{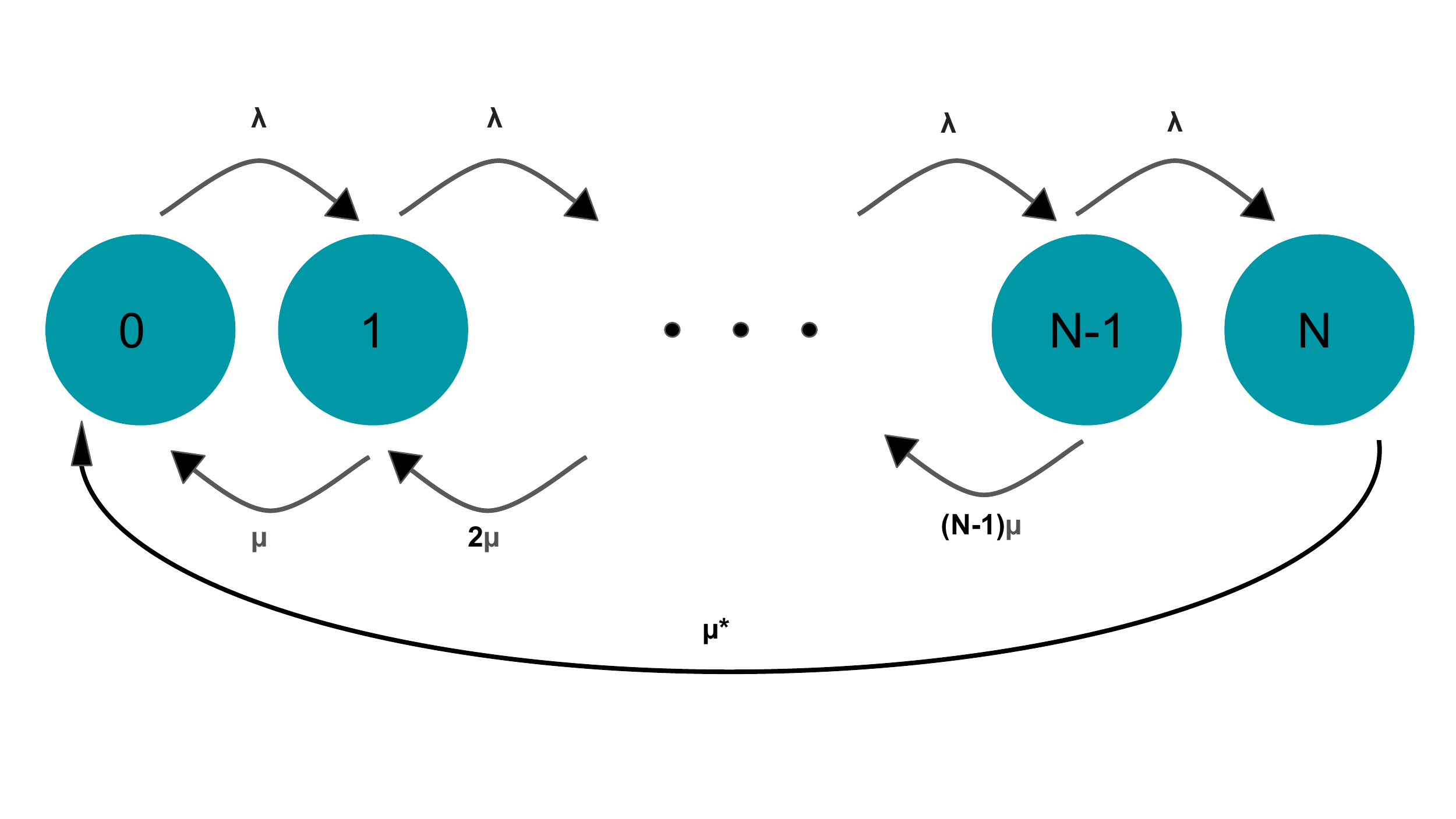}
\end{center}
\caption{Circular Markov chain model with $N+1$ states corresponding to the number of particles in the system.  The arrows indicate the possible interstate
  transitions and their associated rates.}\label{fig:markovchain}
\end{figure}

Particles enter an initially empty channel with a finite capacity according to a Poisson process with intensity $\lambda$ . 
The channel remains open 
if it contains fewer than $N$ particles. 
In the open state, any particle that is present exits the channel at a rate $\mu$, 
independent of the time already spent inside.
When $N$ particles are present the channel is blocked and, all newly arriving
particles are rejected. After a finite duration of time the blockage is released, and all $N$ particles are simultaneously ejected, at a rate 
$\mu^*< \mu$. See Fig. \ref{fig:markovchain}.
The number of particles in the system is a Markov process with state space $S=\{0,1,\ldots,N\}$ 
and infinitesimal generator
\begin{equation}
\bf{Q_N} = \left(\begin{array}{ccccc}
-\lambda&\lambda&&&0\\
\mu&-(\lambda+\mu)&\lambda&&\\
&\ddots&\ddots&\ddots&\\
&&(N-1)\mu&-(\lambda+(N-1)\mu)&\lambda\\
\mu^*&&&0&-\mu^*
\end{array}
\right).
\end{equation}
The system evolves according to the forward Kolmogorov equation
\begin{equation}
 \frac{d\bf{P_N}}{dt}=\bf{P_N}Q_N.
\end{equation}
where $\bf{P_N}$ denotes the state vector $[\pi_0(t),\pi_1(t),...\pi_n(t)]$ with $\pi_k(t)$ being the probability that the system is in state
$k$ at time $t$.

The system is an example of a circular Markov processes, in that all but the Nth state can only transition to a neighboring state. 
The steady-state probabilities of the circular processes are explicitly calculated in Adan and Resing \cite{Adan2002}. 
In the steady state, i.e. $\bf{\dot{P}}_N =0$, the probabilities are given by $\pi_N=C_N \lambda^N$
and, for $k=1,2,\ldots,N$,
\begin{equation}
\pi_{N-k} = C_N \mu^* \sum_{j=1}^k\left(\lambda^{N-j} \mu^{j-1} \prod_{i=1}^{j-1} [N-k+i]\right),
\end{equation}
where $C_N$ is a normalization constant so that $\sum_{i=0}^N \pi_i = 1$. The explicit formula is
\begin{equation}\label{eq:cn}
 C_N = \left(\lambda^N+\mu^*\sum_{j=0}^{N-1}\frac{N!}{(j+1)(N-j-1)!}\mu^j\lambda^{N-1-j}\right)^{-1}.
\end{equation}

Following the analysis presented by Cohen \cite{CohenSingle1982} we find the following formula for the mean first passage time from the empty state 0 to the blocked state
with $N$ particles
\begin{equation}\label{eq:mfpt}
\nu_{0,N} = \frac{1}{\lambda} \sum_{m=0}^{N-1} m! \sum_{k=0}^m \frac{1}{k!}\left(\frac{\lambda}{\mu}\right)^{k-m}.
\end{equation}
(see Appendix).

\begin{figure}
\begin{center}
\includegraphics[width=8cm]{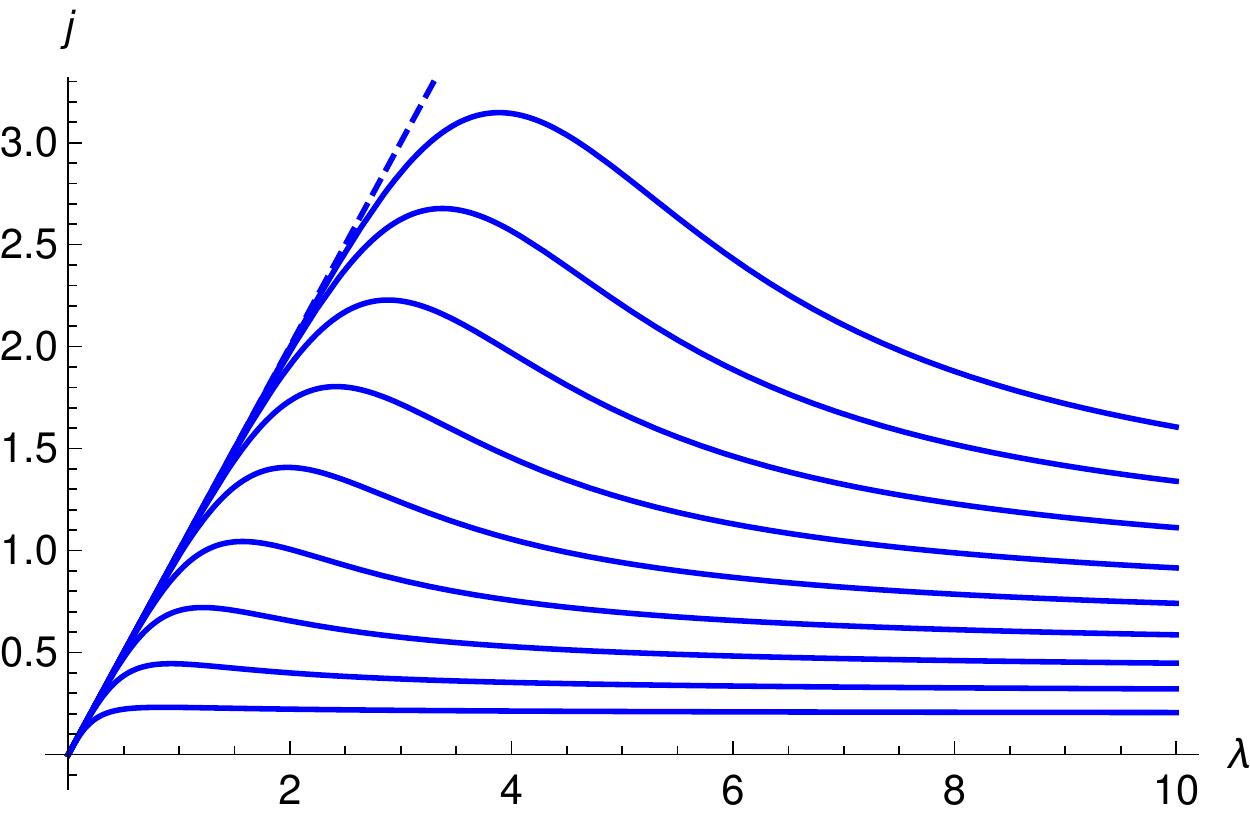}
\end{center}
\caption{Steady state throughput, $j$, Eq. (\ref{eq:thruputss}), as a function of the intensity, $\lambda$, for $\mu=1, \mu*=0.1$ and $N=2,3,...,10$ bottom to top. 
The dashed line shows the behavior in the limit of small intensity, $j=\lambda$. The limiting value at large intensity is given by $N\mu^*$.}\label{fig:thruN}
\end{figure}

The throughput can be calculated by noting that state $k,\;0< k <N$, contributes $k \mu \pi_k(t)$ particles per unit time, which is 
the rate of exiting particles $k\mu$ times the probability that the channel is in state $k$.  There is an additional contribution 
from the blocked state whose $N$ particles 
are simultaneously ejected at the rate $\mu^*$, which gives a contribution of $N \mu ^* \pi_N(t)$.
The general expression of the time dependent throughput is thus the sum of these contributions
\begin{equation}\label{eq:thruput}
j_N(\lambda,t) = \left( \sum_{k=1}^{N-1} k \mu \pi_k(t) \right) + N \mu ^* \pi_N(t) .  
\end{equation} 
In the steady state the rate of exiting particles also corresponds to the incident particulate flux minus the part that is rejected when the system is in the blocked state:
\begin{equation}\label{eq:thruputss}
j_N(\lambda) = \lambda(1-\pi_N).
\end{equation} 
From $\pi_N=C_N\lambda^N$ and Eq. (\ref{eq:cn}) we find the following explicit expression 
\begin{equation}
 j_N(\lambda)=\frac{\lambda}{\left[\mu^*\sum_{j=0}^{N-1}\frac{N!}{(j+1)(N-j-1)!}\mu^j\lambda^{-(j+1)}\right]^{-1}+1}.
\end{equation}
We confirm the expected result that, in the large intensity limit
\begin{equation}
 j_N(\lambda\rightarrow\infty) = N\mu^*.
\end{equation}
which corresponds to the situation where no particle can cross the channel without blockage. The flux is given by the number of particles
trapped in each blockage, $N$,  times the rate of the channel release, $\mu^*$.
At low intensity the throughput is given by
\begin{equation}
 j_N(\lambda)=\lambda - \frac{\lambda^{N+1}}{(N-1)!\mu^{N-1}\mu^*}+O(\lambda^{N+2}).
\end{equation}
In this limit almost all particles cross the channel, which corresponds to the leading term $\lambda$. The 
decrease of the flux corresponding to the term of order $\lambda^{N+1}$ is due to the rare events where the channel is blocked and is given by $\frac{\lambda}{\mu^*\tau}$, where $\tau$ is the mean time of blockage at low intensity\cite{Barre2015a}, $\tau=(N-1)!\frac{\mu^{N-1}}{\lambda^N}$.
The outgoing flux as a function of the intensity is shown for several values of $N$ and for $\mu=1,\mu^*=0.1$  in Fig. \ref{fig:thruN}. 
One notes the initial linear regime corresponding to no loss of the incoming flux, 
which increases in importance with increasing $N$. This is followed by a maximum throughput at a finite intensity. 

Let us compare our model with the well-known $M/M/N/N$ queue \cite{Adan2002}, i.e., one  with exponentially distributed arrival and service times with rates $\lambda$ and $\mu$, respectively, $N$ servers and no waiting line \cite{medhi2002stochastic}. 
The Markov chain of this model is similar to Fig. \ref{fig:markovchain} except that there is no direct transition from state $N$ to state 0. Instead there is a transition from 
state $N$ to state $N-1$ with rate $N\mu$.
An arriving unit is lost to the system
if all $N$ servers are busy. The probability of this event is given by the Erlang loss formula
\begin{equation}
 \pi_N=\frac{\frac{1}{N!}(\frac{\lambda}{\mu})^N}{\sum_{i=0}^N\frac{1}{i!}(\frac{\lambda}{\mu})^i}
\end{equation}
The throughput, given by substituting this probability in Eq. (\ref{eq:thruputss}), always increases monotonically towards the maximum value $N\mu$ as $\lambda$ increases.

In the following we examine in more detail the behavior of the systems $N=2,3$. 
For simplicity, and without loss of generality, we will often set $\mu=1$. This is equivalent to taking the unit of time as $\mu^{-1}$.

\subsection{N=2}
The explicit equations describing the evolution of the three state probabilities, with $\mu = 1$, are;
\begin{eqnarray}\label{eq:master2}
\dot{\pi_0}&=-\lambda\pi_0(t)+\mu\pi_1(t)+\mu^*\pi_2(t),\nonumber\\
\dot{\pi_1}&=\lambda\pi_0(t)-(\mu+\lambda)\pi_1(t),\nonumber\\
\dot{\pi_2}&=\lambda\pi_1(t)-\mu^*\pi_2(t).
\end{eqnarray}
and the time dependent throughput is
\begin{equation}\label{eq:j2det}
j(\lambda,t) = \mu \pi_1(t) + 2 \mu^* \pi_2(t).
\end{equation} 
As the system evolves towards a steady state the throughput approaches a constant value.

\begin{figure}
\begin{center}
\includegraphics[width=8cm]{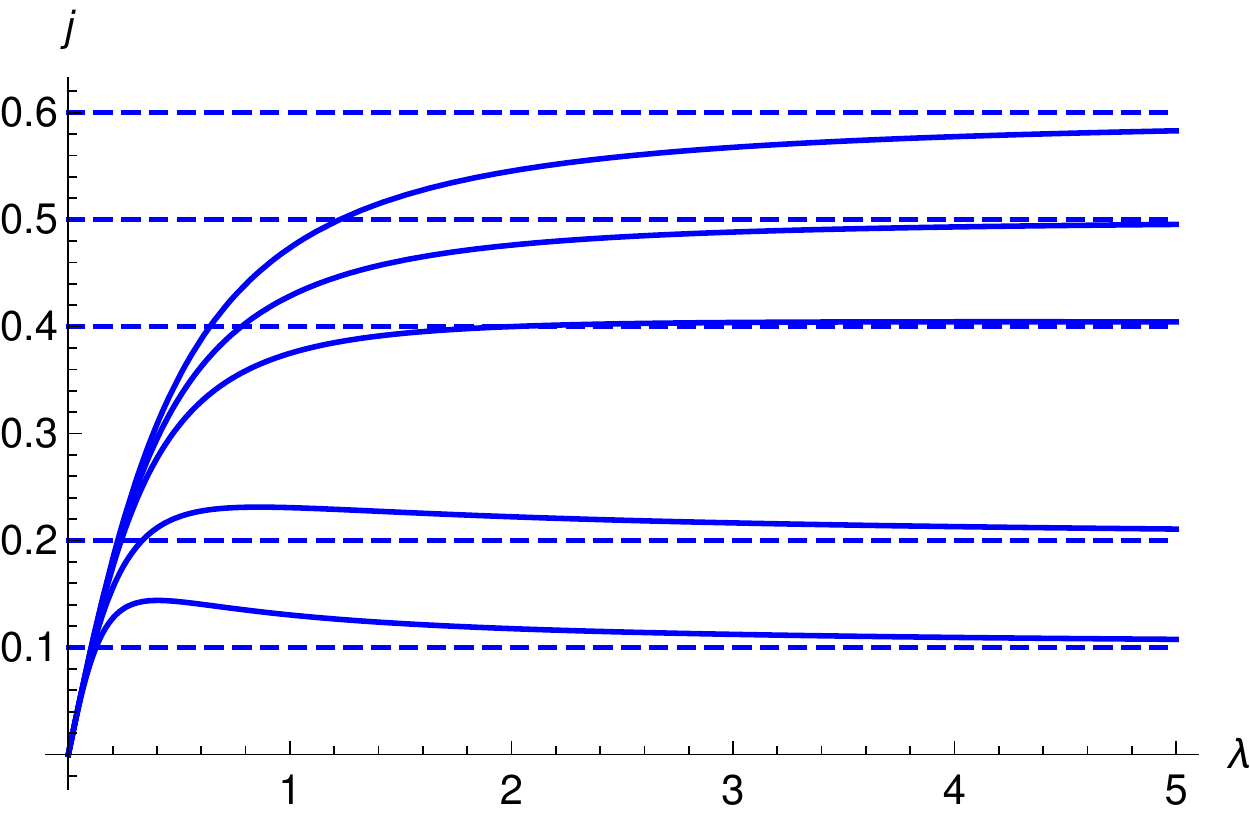}
\end{center}
\caption{$N=2$. Steady state throughput, $j$ as a function of intensity, $\lambda$ for $\mu=1,\mu*=0.3,0.25,0.2,0.1,0.05$ top to bottom. The dashed lines show the limiting value, $2\mu^{*}$. Global maxima are evident for certain values of $\mu^*$ (see text).}\label{fig:thru}
\end{figure}

{\bf Steady State Behavior.} In the steady state, the stationary probabilities are 

\begin{equation}
 [\pi_0,\pi_1,\pi_2]=C_2[(\lambda+\mu)\mu^*,\lambda\mu^*,\lambda^2].
\end{equation}
with $C_2=((2\lambda+\mu)\mu^*+\lambda^2)^{-1}$.
As expected Eq. (\ref{eq:thruput}) and Eq. (\ref{eq:thruputss}) yield the same result for the 
steady state throughput:

\begin{equation}\label{eq:omega2}
 j(\lambda)=\lambda\frac{2\lambda+\mu}{\lambda^2/\mu^{*}+2\lambda+\mu}
 \end{equation}
Figure \ref{fig:thru}  displays the steady state throughput as a function of the intensity, $\lambda$.
The limiting values of the steady state throughput at small and large intensity are
\begin{eqnarray}
 j(\lambda)=\lambda-\frac{\lambda^3}{\mu\mu^*}+O(\lambda^4),\\
 \lim_{\lambda\rightarrow\infty}j(\lambda)=2\mu^*.
\end{eqnarray}

\begin{figure}
\begin{center}
\includegraphics[width=8cm]{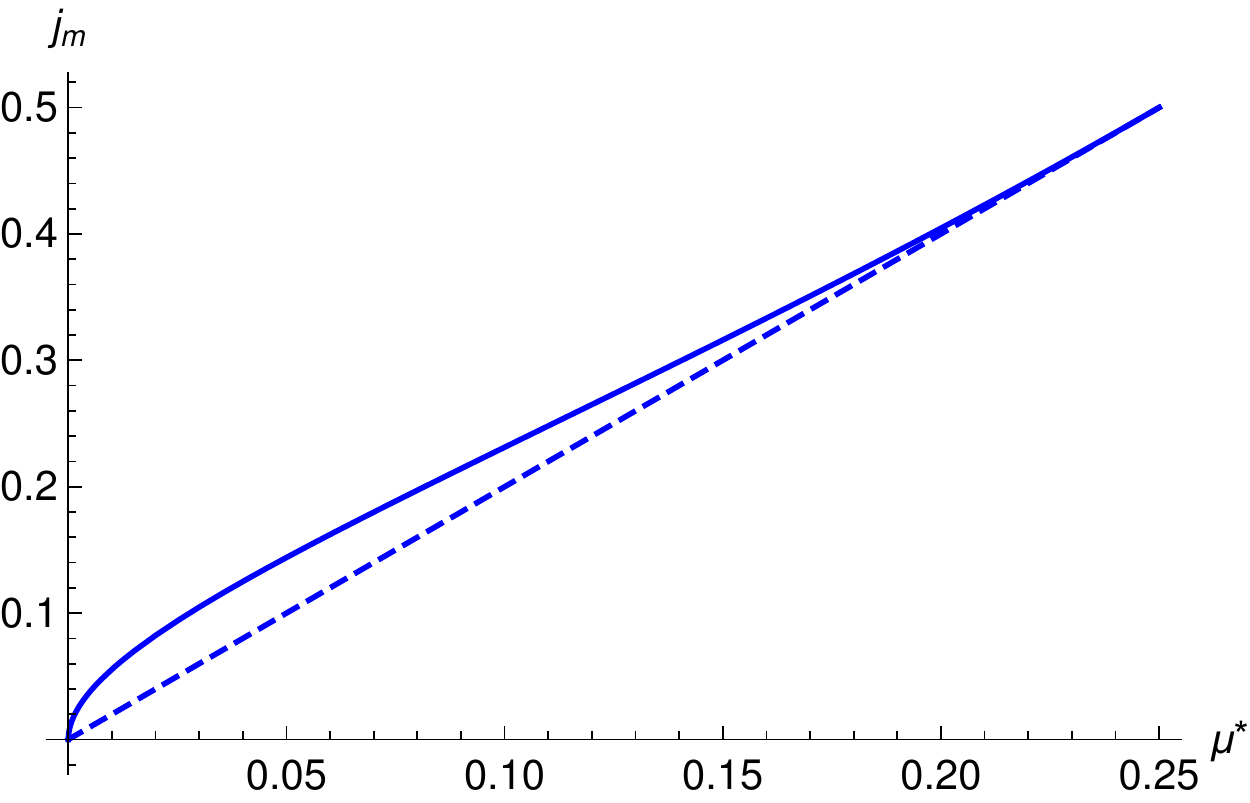}
\end{center}
\caption{$N=2$. Maximum throughput, $j_{\rm max}$, as a function of $\mu^*$ ($\mu=1$). The dashed lines shows the limiting value, $2\mu^{*}$.}\label{fig:maxthru}
\end{figure}

We search for a non-trivial maximum throughput at finite $\lambda$ by seeking solutions of
$dj/d\lambda=0$ which requires $4\lambda\mu\mu^*+\mu^2\mu^*-\lambda^2(\mu-4\mu^*)=0$. The solutions can be written as
\begin{equation}\label{eq:lmaxss}
 \frac{\lambda}{\mu}=\frac{\sqrt{r}+2r}{1-4r}\ge 0,
\end{equation}
where $r={\mu^*}/\mu$. Thus, solutions exist for $r<r_c=1/4$. If $r>1/4$ there is no maximum at finite $\lambda$. If $r<1/4$ the maximum
throughput is given by 
\begin{equation}
 j_{\rm max}=\frac{\mu\sqrt{r}}{2(1-\sqrt{r})}=\frac{\mu^*}{2(1-\sqrt{r})\sqrt{r}}
\end{equation}
Figure \ref{fig:maxthru} shows that this approaches the limiting value $2\mu^*$ as $\mu⁸\rightarrow 1/4$.

{\bf Mean Time to First Blockage.}  From Eq. (\ref{eq:mfpt}) we have 
\begin{equation}
 \nu_{0,2}=\frac{2}{\lambda}+\frac{\mu}{\lambda^2}.
\end{equation}
This provides an alternative route to the probability that the system is in the blocked state, $\pi_2$, and thus the steady state throughput. 
Following \cite{BTV2013} we note that in the steady state there is an alternation of open and blocked states with average durations $\nu_{0,2}$ and 
$1/\mu^*$, respectively. Thus $\pi_2=(1/\mu^*)/((1/\mu^*)+\nu_{0,2}) =\lambda^2/((\mu+2\lambda)\mu^*+\lambda^2)$, as obtained previously.

\begin{figure}
\begin{center}
\includegraphics[width=8cm]{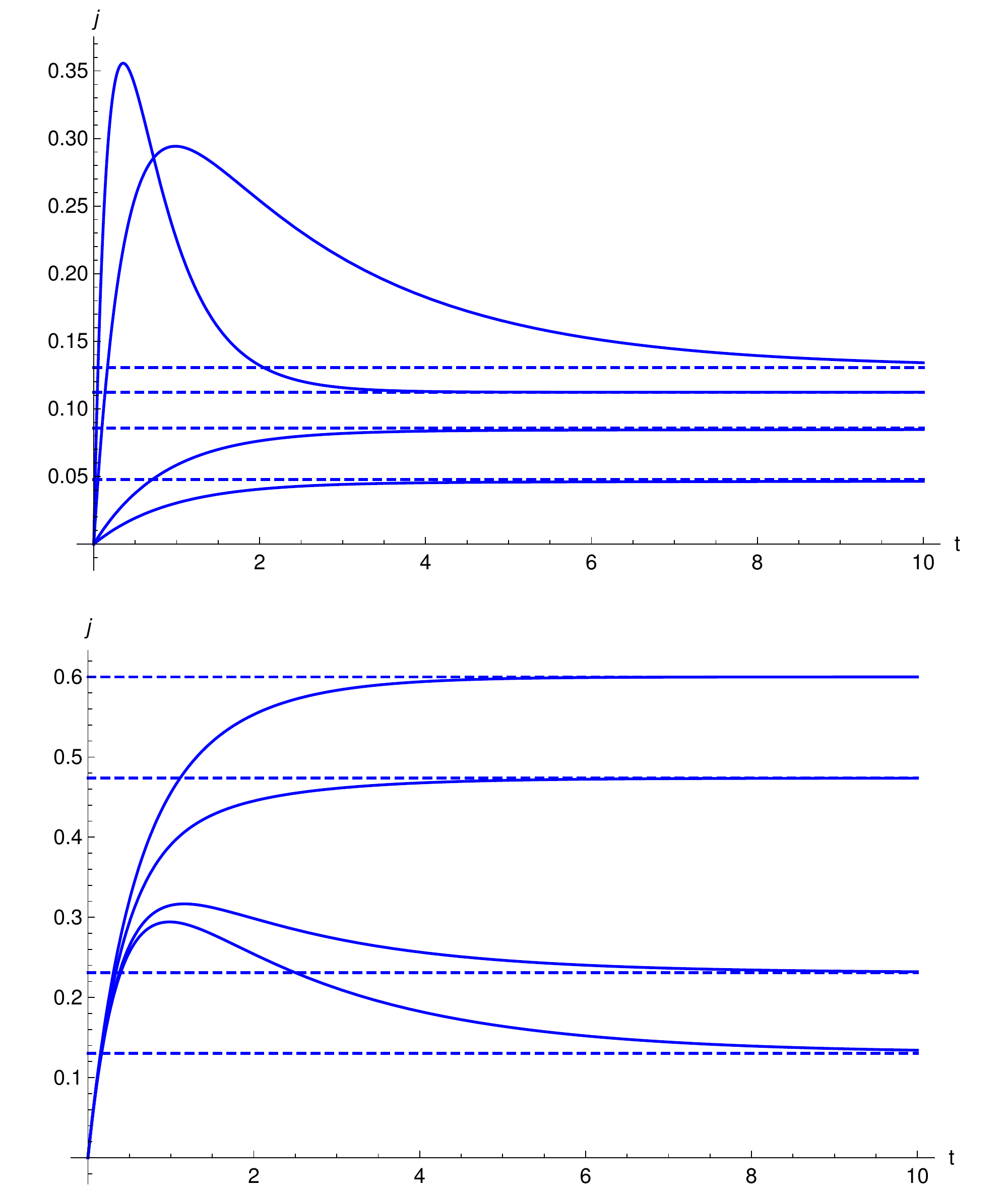}
\end{center}
\caption{Time evolution of $j$ for $N=2$. Top: $ \mu^*= 0.05$, $\lambda$ = 3, 1, 0.1, 0.05, top to bottom. Bottom: $\lambda =1$, $ \mu^* = 0.05, 0.1, 0.3, 0.5$ bottom to top.
The dashed lines show the steady state value, Eq. (\ref{eq:omega2}). }\label{fig:OmegaTime}
\end{figure}

{\bf Kinetic Behavior.} The system of equations Eq. (\ref{eq:master2}) may be solved analytically.
The eigenvalues of the matrix $Q_3$ are $0$ (associated with the conservation of the total probability), and two real negative values $\gamma_{1,2}$
given by
\begin{equation}
 \gamma_{1,2}=-\frac{\mu+\mu^*+2\lambda\pm\beta}{2}
\end{equation}
where $\beta =\sqrt{(\mu- \mu^*)^2 +4\lambda(\mu-\mu^* ) }$.
The probabilities are given by
\begin{equation}
 \pi_i(t)=\pi_i+a_i e^{\gamma_1 t}+b_i e^{\gamma_2 t}
\end{equation}
where $\pi$ are the stationary values and $a_i$ and $b_i$ are determined by the initial conditions for $i=0,1,2$.
One easily obtains that
\begin{eqnarray}
 a_0&=\frac{\gamma_2}{\beta}(1-\pi_0)+\frac{\lambda}{\beta},&b_0=-\frac{\gamma_1}{\beta}(1-\pi_0)-\frac{\lambda}{\beta}\\
 a_1&=-\frac{\gamma_2}{\beta}\pi_1-\frac{\lambda}{\beta},&b_1=\frac{\gamma_1}{\beta}\pi_1+\frac{\lambda}{\beta}\\
 a_2&=-\frac{\gamma_2}{\beta}\pi_2,&b_2=\frac{\gamma_1}{\beta}\pi_2.
\end{eqnarray}
Figure \ref{fig:OmegaTime} illustrates the behavior of the time dependent throughput for different parameter values. We note
two distinct behaviors: either the throughput increases monotonically to the steady state value, or it displays a maximum at a finite time before decreasing to
the steady state value.

\begin{figure}
\begin{center}
\includegraphics[width=8cm]{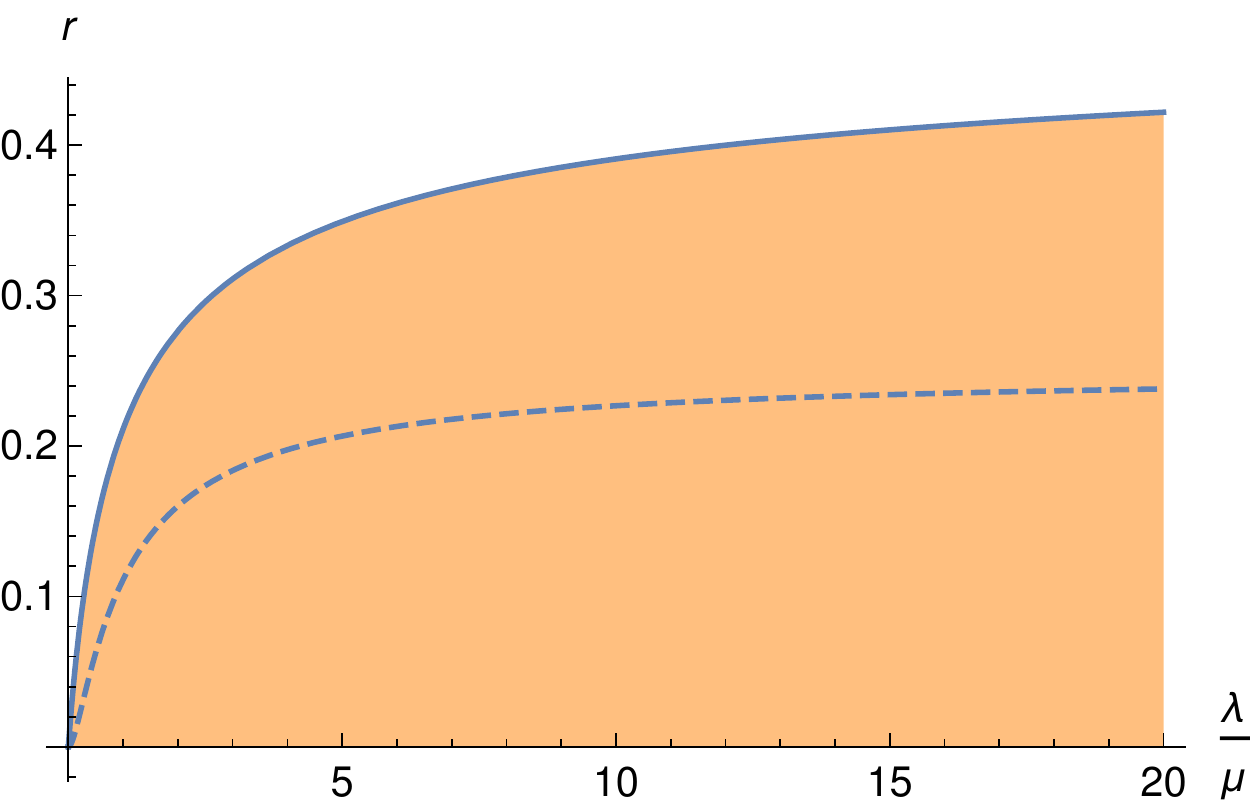}
\end{center}
\caption{State diagram for $N=2$. The region below the upper line, Eq. (\ref{eq:kinetboundary}), corresponds to the parameter space in which a maximum throughput occurs at a finite time given by Eq. (\ref{eq:tmax}). 
No maximum occurs for any value of $\lambda$ if $r>1/2$.
The lower, dashed line, Eq. (\ref{eq:lmaxss}) corresponds to the intensity that maximizes the steady state throughput for a given value of $r=\mu^*/\mu$. No maximum occurs for finite $\lambda$ if $r>1/4$.}\label{fig:KPhase}
\end{figure}

By using Eq.(\ref{eq:j2det}), one obtains
\begin{equation}
\frac{\partial j(t)}{\partial t}=\gamma_1 (\mu a_1 +2\mu^* a_2) e^{\gamma_1 t}+\gamma_2 (\mu b_1 +2\mu^* b_2)  e^{\gamma_2 t}
\end{equation}
The solution for $\frac{\partial j(t)}{\partial t}=0$ is given by
\begin{equation}\label{eq:tmax}
  t_{\rm max} = \frac{1}{\beta}\ln\left(\frac{\mu^*(\mu+2\lambda)+\gamma_1\mu }{\mu^*(\mu+2\lambda)+\gamma_2\mu}\right).
\end{equation}
Note that the denominator of Eq.\ref{eq:tmax} vanishes when
\begin{equation}\label{eq:kinetboundary}
\mu^*_b(\lambda)=\frac{\mu}{2}\left(1 - \sqrt{\frac{\mu}{2\lambda+\mu}}\right),
\end{equation}
which provides the boundary of the set of parameter values at which $j(t)$ displays a maximum.
It is the limiting value of $\mu^*$, for a given value of $\lambda$, at which the nontrivial maxima of the time dependent throughput exists.
We note that if $r>1/2$ no maximum at finite time exists; rather the throughput approaches
the steady state value from below.  The kinetic and steady state behaviors are shown in Fig. \ref{fig:KPhase}.

\subsection{N=3}

The master equations describing the evolution of the four state probabilities are;
\begin{eqnarray}\label{eq:master3}
\dot{\pi_0}&=-\lambda\pi_0(t)+\mu\pi_1(t)+\mu^*\pi_3(t),\nonumber\\
\dot{\pi_1}&=\lambda\pi_0(t)-(\mu+\lambda)\pi_1(t) + 2\mu\pi_2(t),\nonumber\\
\dot{\pi_2}&=\lambda\pi_1(t)-(2\mu+\lambda)\pi_2(t),\nonumber\\
\dot{\pi_3}&=\lambda\pi_2(t)-\mu^*\pi_3(t).
\end{eqnarray}
And the time dependent throughput is;
\begin{equation}
j(\lambda,t) = \mu\pi_1(t) + 2\mu \pi_2(t) + 3 \mu^* \pi_3(t).
\end{equation} 

{\bf Steady State Behavior.} The steady state probabilities are
\begin{equation}
 [\pi_0,\pi_1,\pi_2,\pi_3]=C_3[(\lambda^2+\lambda\mu+2\mu^2)\mu^*,\lambda(\lambda+2\mu)\mu^*,\lambda^2\mu^*,\lambda^3].
\end{equation}
with $C_3=(\lambda^3+3\mu^*\lambda^2+3\mu\mu^*\lambda+2\mu^2\mu^*)^{-1}$.
And the throughput is
\begin{equation}\label{eq:}
 j(\lambda)=\lambda(1-\pi_3)=\frac{\lambda(3\lambda^2+3\lambda\mu+2\mu^2)\mu^*}{\lambda^3+3\mu^*\lambda^2+3\mu\mu^*\lambda+2\mu^2\mu^*}
\end{equation}
with the limiting values at small and large intensities
\begin{eqnarray}
 j(\lambda)&=\lambda-\frac{\lambda^4}{2\mu^2\mu^*}+O(\lambda^5), \\\
 \lim_{\lambda\rightarrow\infty}j(\lambda)&=3\mu^*,
\end{eqnarray}
respectively.

\begin{figure}
\begin{center}
\includegraphics[width=8cm]{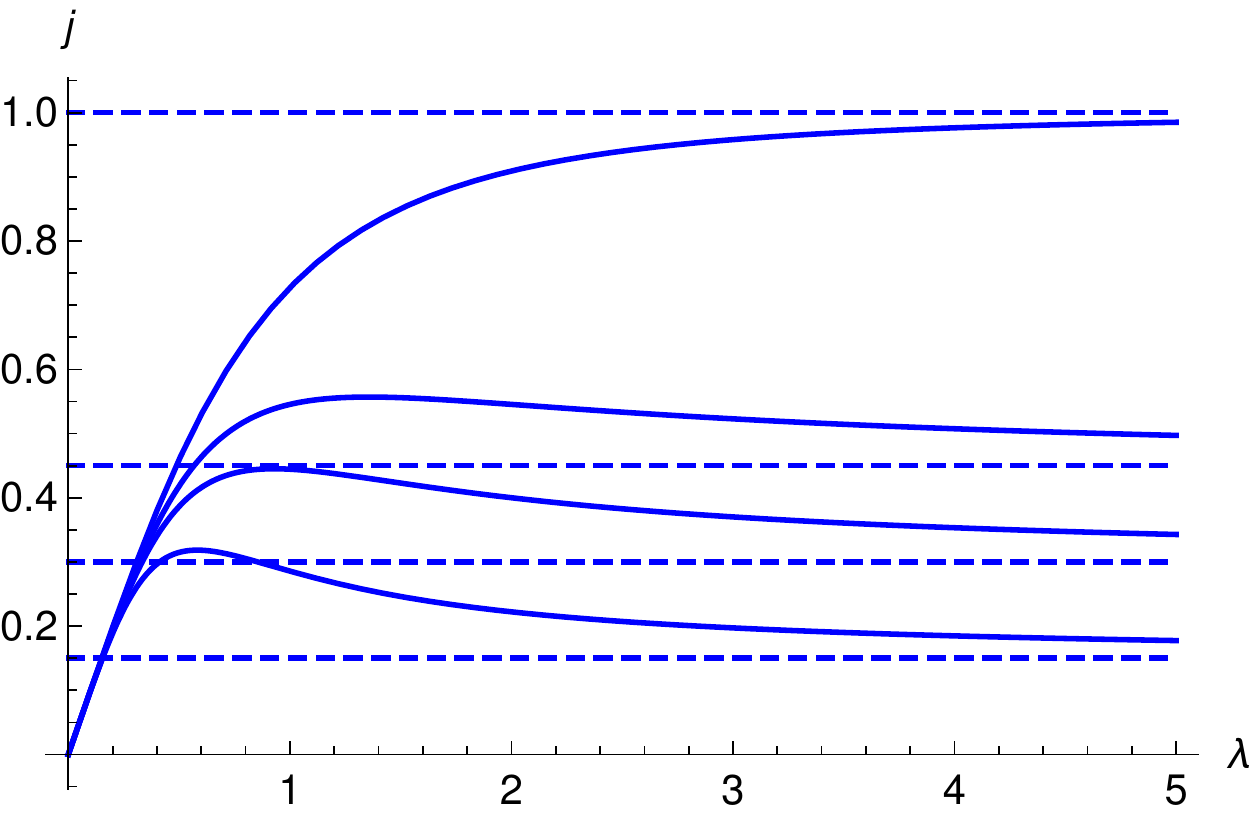}
\end{center}
\caption{$N=3$. Steady state throughput, $j$, as a function of $\lambda$ for $\mu=1,\mu*=0.3333,0.15,0.1,0.05$ top to bottom. The dashed lines show the limiting value, $3\mu^{*}$.}\label{fig:thru3}
\end{figure}

Some examples are shown in Fig. \ref{fig:thru3}. For $r<r_c=1/3$ there is a maximum throughput at finite intensity. 

\begin{figure}
\begin{center}
\includegraphics[width=8cm]{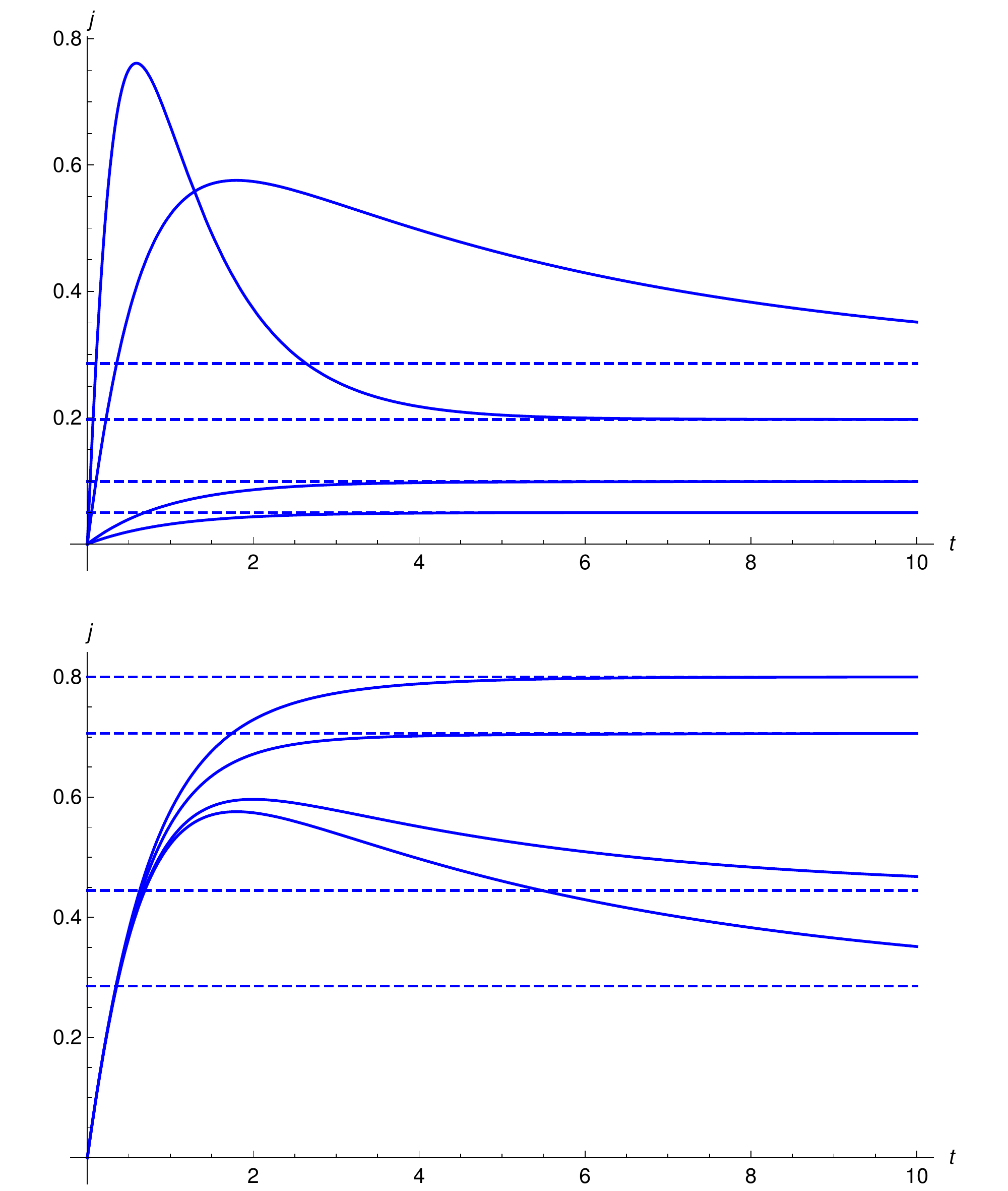}
\end{center}
\caption{$N=3$. Time evolution of $j(t)$. Top: $ \mu^*= 0.05$ fixed, for $\lambda = 3,1,0.1,0.05,$ top to bottom. Bottom: $\lambda =1$ fixed; $ \mu^* = 0.05,.1,0.3,0.5,$ bottom to top.  The dashed lines show the 
steady state value, Eq. (\ref{eq:}).}\label{fig:OmegaTime3}
\end{figure}

{\bf Kinetic Behavior.} To explore the transient behavior, we solved Eqs. (\ref{eq:master3}) numerically. Some results are shown in Fig. \ref{fig:OmegaTime3}. 
The same two distinct behaviors, as remarked for $N=2$, are present. No maximum at finite time is observed if $\mu^* > 0.75$.

\subsection{Comparison with Reversible Model with Constant Transit and Blockage Times}
The model introduced above is fully stochastic in the sense that the interval between particle entries, transit and blockage times are all sampled from exponential distributions
with given rates. While this choice can lead to a considerable simplification of the mathematics, it may be unrealistic in certain physical applications. A particle cannot traverse
a channel in zero time, while according to the exponential distribution this is the most likely outcome. It therefore seems reasonable to consider alternative models with 
constant transit and/or blockage times \cite{BTV2013}.

For $N=1$ (that is the channel is blocked by the entry of the first particle) it is easy to show that the steady state properties for the model with an exponentially distributed blocking time
with rate $\mu^*$ are the same as for a model with a constant blocking time, $\tau_b$, if the deblocking rate is equal to the inverse of the (constant) blocking time. 
For example the probabilities that the 
channel is open are  $p_o=\mu^*/(\mu^*+\lambda)$ and $p_o=1/(1+\lambda\tau_b)$, respectively. These are the same if $\tau_b=1/\mu^*$. The kinetics of the two models
are, however, different (For example, when the blockage time is constant the exiting flux is strictly zero for $\tau<\tau_b$, while for an exponentially distributed
blockage time the mean flux is finite for all $t>0$.
We examined the corresponding ($N=2$) model in \cite{BTV2013}, i.e., one with a constant transit and blockage times
$\tau$ and $\tau_b$, respectively. In the steady state the exiting flux is given by
\begin{equation}\label{eq:jss}
j_{\infty}=\frac{\lambda(2-e^{-\lambda\tau})}{\lambda\tau_b(1-e^{-\lambda\tau})+2-e^{-\lambda\tau}}.
\end{equation}
Comparing this to the throughput of the first model above with $N=2$,
Eq. (\ref{eq:omega2}), we see that there is
no simple mapping when one replaces $\tau$ with $1/\mu$ and $\tau_b$
with $1/\mu^*$. If, however, we introduce an effective exit (service) rate we can map the two systems:
\begin{eqnarray}
 \mu^* &= \frac{1}{\tau_b},\nonumber\\
 \mu &= \frac{\lambda e^{-\lambda\tau}}{1-e^{-\lambda\tau}}.
\end{eqnarray}
That is, by substituting these equations in Eq. (\ref{eq:omega2}) we obtain Eq. (\ref{eq:jss}).
Thus, we expect the steady state behavior to be qualitatively similar. In particular
the exiting flux, Eq. (\ref{eq:jss})  may be
maximized for a finite value of the intensity if $\tau/\tau_b<0.16$.

\section{Irreversible Blockage Model with a Source of Finite Duration}
\label{sec:irrevmodel}

This model, originally introduced by Gabrielli et al. \cite{Gabrielli2013}, considers a stream of particles entering a channel according to a Poisson process of intensity $\lambda$ 
with the channel carrying capacity set to $N=2$. A single particle exits in a time $\tau$, but if ever two 
particles are present the channel blocks irreversibly. Properties of interest, including the survival probability at time $t$, mean blockage time, exiting particle flux, and the total number of 
exiting particles can be calculated exactly \cite{Gabrielli2013,Talbot2015}. Here we consider the problem of maximizing the total number of exiting particles in a given finite time. 

The average number of particles that exit in the time interval $(0,t_s)$ can be computed by integrating the exiting particle flux,
\begin{equation}\label{eq:mj}
 m(t_s)=\int_1^{t_s}j(t)dt
\end{equation}
We take the lower limit of the integral $t=\tau=1$, as no particle can exit before this time (assuming that the channel is empty at $t=0$.)
For $t_s\rightarrow\infty$ we can show that
\begin{equation}\label{eq:mbarlim}
 m(\infty) = \tilde{j}(u=0) = \frac{1}{e^\lambda-1}
\end{equation}
where $\tilde{j}(u)=\int_0^{\infty}e^{-ut}j(t)dt$ is the Laplace transform. 
The number of exiting particles tends to infinity as $\lambda\rightarrow 0$ and to zero as $\lambda\rightarrow\infty$. 
More interesting, however, is the situation for finite $t_s$. In this case there is clearly a finite entering intensity that optimizes the total number of
exiting particles. If the intensity is too small, blocking is unlikely but few particles enter, while if $\lambda$ is too large more particles are injected
but blocking is highly probable.
 
\begin{figure}
\begin{center}
\includegraphics[width=8cm]{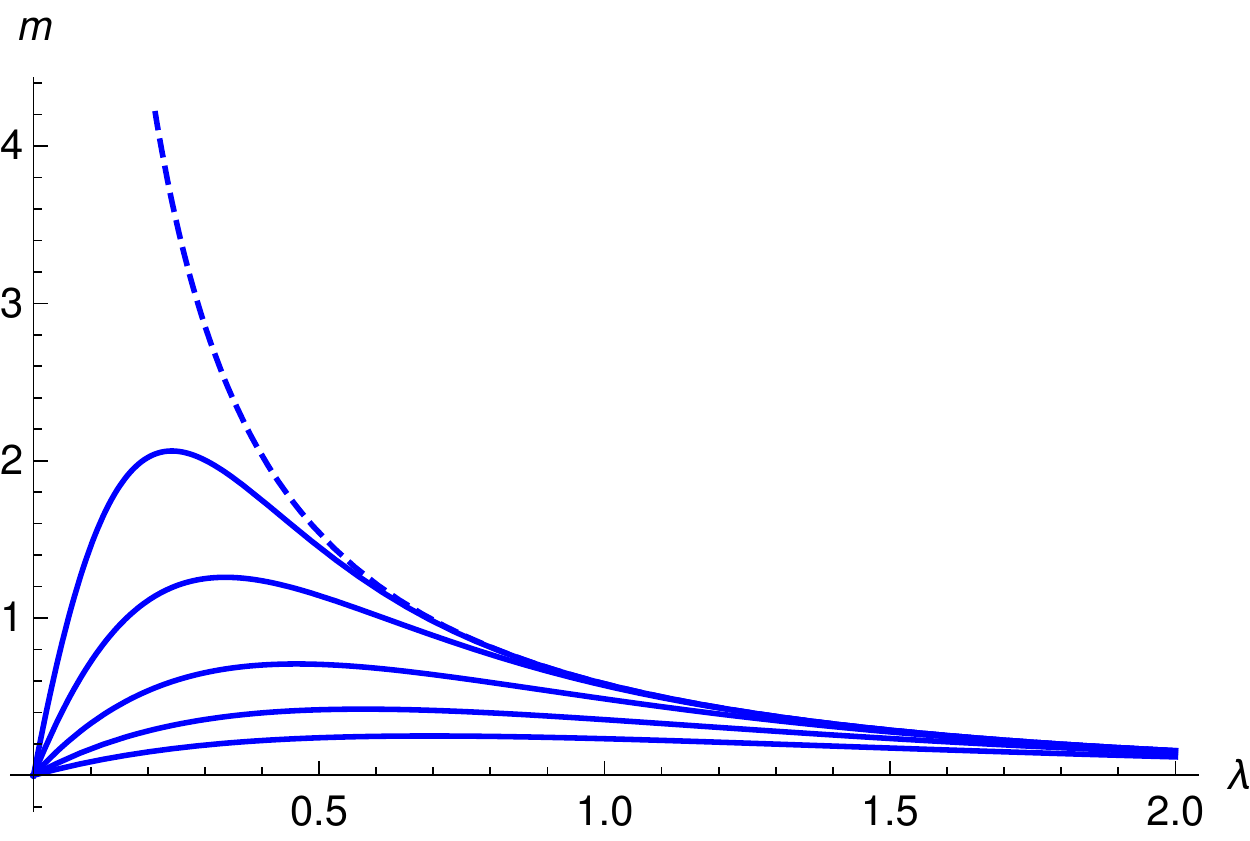}
\end{center}
\caption{Average number of exiting particles as a function of the intensity for different stopping times, $t_s=32,16,8,4,2,1.5$ top to bottom. The dashed line shows the long-time
limit, Eq. (\ref{eq:mbarlim}).}\label{fig:mout}
\end{figure}

The explicit equation for the flux at time $t$ \cite{Talbot2015} is given by 
\begin{equation}
 j(t) = e^{-\lambda t}\sum_{k=1}^{\left \lfloor{t}\right \rfloor}\frac{\lambda^k (t-k)^{k-1}}{(k-1)!}.
\end{equation}
Substituting this in Eq. (\ref{eq:mj}) gives 
\begin{equation}
\fl m(t_s)=\sum_{k=1}^{\left \lfloor{t_s}\right \rfloor}\exp(-\lambda k)\left[1-\frac{\Gamma(k,\lambda(t_s-k))}{(k-1)!}\right]
 =\sum_{k=1}^{\left \lfloor{t_s}\right \rfloor}\exp(-\lambda k)\frac{\gamma(k,\lambda(t_s-k))}{(k-1)!},
\end{equation}
where $\Gamma(k,x)$ and $\gamma(k,x)$ are the upper and lower incomplete gamma functions, respectively. Some results are shown in Fig. \ref{fig:mout}. 
It can be seen that the expected number of exiting particles displays a maximum at finite intensity for a finite stopping time, $t_s$. We observe that the 
value of $\lambda$ that maximizes the output increases as $t_s$ decreases and that the maximum sharpens as $t_s$ increases.

The explicit expression for the intensity that maximizes the output is a piece-wise function. For $1\le t_s\le 2$, 
\begin{equation}\label{eq:mt1}
 m(t_s) = e^{-\lambda}-e^{-\lambda t_s}.
\end{equation}
It is easy to show that a maximum occurs for
\begin{equation}
 \lambda = \frac{\ln(t_s)}{t_s-1}
\end{equation}
which gives $\lambda=\ln(2)$ when $t_s=2$.

For $2\le t_s\le 3$
\begin{equation}
 m(t_s) = e^{-\lambda}-e^{-\lambda t_s}+e^{-2\lambda}-e^{-\lambda t_s}(1+\lambda(t_s-2)).
\end{equation}
$m(t)$ being a  piece-wise continuous function, one easily checks that for $t_s=2$, the value is the same as that given by Eq. (\ref{eq:mt1}).
There is no analytic expression of the value of $\lambda$ that maximizes this expression for $t>2$, but 
numerical solutions are, however, straightforward. Moreover, for large $t_s$, which corresponds to small intensity $\lambda$, one can obtain an asymptotic solution. By using the Laplace's method\cite{Talbot2015}, $j(t)$ decays as 
\begin{equation}\label{eq:jasympt}
 j(t)\simeq \lambda e^{-\lambda^2 t}.
\end{equation}
Integrating Eq. (\ref{eq:jasympt}), one obtains that $m(t_s)$ is given by
\begin{equation}\label{eq:masympt}
 m(t_s)\simeq \frac{1}{\lambda} (1-e^{-\lambda^2 t_s}).
\end{equation}
Differentiating Eq. (\ref{eq:masympt}) with respect of $\lambda$, one obtains a maximum of $m(t_s)$
when 
\begin{equation}
 \lambda \simeq \frac{C}{\sqrt{t_s}}.
\end{equation}
where $C$ is a constant.

The limitation of the flux route is that it does not allow us to calculate higher moments of the exiting particle distribution, but only the mean value. To go further, we therefore introduce
the function $f(m,t)$ giving the probability that $m$ particles have exited at time $t$, regardless of the state of the channel (open or closed) at time $t$
(the joint probabilities that $m$ particles have exited at time $t$ and the channel is still open are discussed in the Appendix). 
The equations describing the evolution of these functions are as follows:
\begin{eqnarray}
 \frac{df(0,t)}{dt}&=-\lambda e^{-\lambda } q_s(0,t-1)\label{eq:f0}\\
  \frac{df(1,t)}{dt}&=\lambda e^{-\lambda }q_s(0,t-1)-\lambda e^{-2\lambda }q_s(1,t-2)\label{eq:f1}\\
 \frac{df(n,t)}{dt}&=\lambda e^{-2\lambda }q_s(n-1,t-2)-\lambda e^{-2\lambda }q_s(n,t-2),\;n > 1,
\end{eqnarray}
where $q_s(n,t)$ is the joint probability that $n$ particles have entered the channel and the channel still open \cite{Talbot2015}.
The loss term for the evolution of $f(0,t)$ is the result of a particle exiting the channel at time $t$ that had previously entered the channel at $t-1$.
This term is also the gain term for the evolution of $f(1,t)$. 
For $n>1$ the gain term consists of the entry of a particle at $t-1$ that exits at $t$. For this to be possible the channel must have been open at $t-2$ 
with $n-1$ particles entering in the interval $(0,t-2)$, which occurs with probability $q_s(n-1,t-2)$,
and no particle must enter in the intervals $(t-2,t-1)$ and $(t-1,t)$ giving rise to the factor of $e^{-2\lambda}$. The loss 
term is similar except that $n$ particles must have entered in the interval $(0,t-2)$ with an additional particle entering at $t-1$ and exiting at $t$. 
The gain term for $n=1$ is slightly different as the particle entering the channel at $t-1$ is the first one so the channel is certainly empty at 
this time. 

We note the telescopic structure of these equations that is consistent with the conservation of probability $\sum_{n=0}^{\infty} df(n,t)/dt=0$.

We can obtain the complete solution by introducing the generating function 
\begin{equation}
 G_f(x,t)=\sum_{n=0}^{\infty}z^n f(n,t).
\end{equation}
Taking the time derivative and substituting the above expressions for $f(n,t)$ we obtain
\begin{equation}
 \frac{\partial G_f(z,t)}{\partial t} = \lambda(z-1)(e^{-\lambda}q_s(0,t-1)+e^{-2\lambda}(G(z,t-2)-q_s(0,t-2))),
\end{equation}
where $G(z,t)=\sum_{n=0}^{\infty}z^n q_s(n,t)$.
Taking the Laplace transform and using the initial condition $G_f(z,0)=1$ we finally obtain
\begin{equation}
 \tilde{G}_f(z,u)=\frac{u+\lambda-\lambda e^{-(u+\lambda)}}{u(u+\lambda-\lambda z e^{-(u+\lambda)})}.
\end{equation}
The individual functions can be recovered from
\begin{equation}
\tilde{f}(n,u)=\frac{1}{n!}\frac{\partial^n\tilde{G_f}(z,u)}{\partial z^n}\biggr|_{z=0}.
\end{equation}
The first two are
\begin{equation}
 f(0,t)=1+(e^{-\lambda t}-e^{-\lambda})\theta(t-1)
\end{equation}
and
\begin{equation}
 f(1,t)=((e^{-\lambda t}(1+\lambda(t-2))-e^{-2\lambda})\theta(t-2)
 -(e^{-\lambda t}-e^{-\lambda})\theta(t-1),
\end{equation}
where $\theta(t)$ is the Heaviside function.
These results can also be obtained by direct solution of Eqs. (\ref{eq:f0}) and (\ref{eq:f1}), respectively.
The first two moments are
\begin{equation}\label{eq:mdeu}
\langle\tilde{m}(u)\rangle=\frac{\partial \tilde{G_f}(u,z)}{\partial z}\biggr|_{z=1}=\frac{\lambda e^{-(u+\lambda)}}{u(u+\lambda-\lambda e^{-(u+\lambda)})},
\end{equation}
and
\begin{eqnarray}
\langle\tilde{m}^2(u)\rangle&=\frac{\partial^2 \tilde{G_f}(u,z)}{\partial z^2}\biggr|_{z=1}+\frac{\partial \tilde{G_f}(u,z)}{\partial z}\biggr|_{z=1}\nonumber\\
&=\frac{\lambda e^{-(u+\lambda)}(\lambda+(u+\lambda) e^{-(u+\lambda)})}{u(\lambda-(u+\lambda) e^{-(u+\lambda)})^2}.
\end{eqnarray}
from which the first and second moments at infinite time may be obtained as
\begin{equation}\label{eq:minf}
 \langle m\rangle = \lim_{u\rightarrow 0}u \langle\tilde{m}(u)\rangle = \frac{1}{e^{\lambda}-1}
\end{equation}
and
\begin{equation}\label{eq:m2u}
 \langle m^2\rangle = \lim_{u\rightarrow 0}u \langle\tilde{m}^2(u)\rangle = \frac{1+e^{\lambda}}{(1-e^{\lambda})^2},
\end{equation}
giving for the variance
\begin{equation}\label{eq:varlimit}
 \langle m^2\rangle - \langle m\rangle^2=\frac{e^{\lambda}}{(e^{\lambda}-1)^2}.
\end{equation}
At small intensity, this behaves as $\lambda^{-2}$, while for large intensity it approaches zero as $e^{-\lambda}$.

In order to obtain the variance of $m$ at time $t$, we have to invert $\langle m^2(u) \rangle$ and $\langle m(u) \rangle$.
By using Eq.(\ref{eq:mdeu}), one has
\begin{eqnarray}
 \langle m(u) \rangle &=\frac{e^{-(u+\lambda)}}{1-e^{-(u+\lambda)}}\left[\frac{1}{u}-\frac{1}{u+\lambda-\lambda e^{-(u+\lambda)}}\right]\\
 &=\sum_{n\geq 1} e^{-n(u+\lambda)}\left[\frac{1}{u}-\sum_{k\geq 0} 
 \frac{ (\lambda e^{-(u+\lambda)})^k}{(u+\lambda)^{k+1}}\right],
\end{eqnarray}
which gives
\begin{equation}
\langle m(t) \rangle= \sum _{n \geq 1} \left(e^{-\lambda  n} \theta (t-n)-e^{-\lambda t} \sum _{k=0} \frac{\lambda ^k (t-k-n)^k \theta (t-k-n)}{k!}\right).
\end{equation}
This expression corresponds to that obtained from the survival probability \cite{BTV2013}. 

Similarly, by using Eq.(\ref{eq:mdeu2}), one obtains
\begin{eqnarray}
 \langle m^2(u) \rangle &=\frac{e^{-(u+\lambda)}(1+e^{-(u+\lambda)})}{(1-e^{-(u+\lambda)})^2 }\left[\frac{1}{u}-\frac{1}{u+\lambda-\lambda e^{-(u+\lambda)}}\right]\nonumber\\
 &- 2\frac{\lambda e^{-2(u+\lambda)}}{1-e^{-(u+\lambda)}}\frac{1}{(u+\lambda-\lambda e^{-(u+\lambda)})^2}.
\end{eqnarray}
By using the identity $\frac{1}{(1-a)^2}=\sum_{n=0}(n+1) a^n$, one has
\begin{eqnarray}
\langle m^2(u) \rangle &=\sum_{n=0}(n+1)(1+e^{-(u+\lambda)})e^{-(n+1)(u+\lambda)} \left[\frac{1}{u}-\sum_{k=0}\frac{(\lambda e^{-(u+\lambda)})^k}{(u+\lambda)^{k+1}}\right]\nonumber\\
&- 2\sum_{n=0}\lambda e^{-(n+2)(u+\lambda)}\sum_{k=0}(k+1)\frac{(\lambda e^{-(u+\lambda)})^k}{(u+\lambda)^{k+1}}.
\end{eqnarray}
that can be inverted to give the second moment as a function of time:
\begin{eqnarray}
\fl\langle m^2(t) \rangle= \sum_{n \geq 0} (1 + n) e^{-(2 + n) \lambda} \left( \theta(t-n-2) + e^\lambda \theta(t-n-1)\right )\nonumber\\
\fl-\sum_{n \geq 0} \sum_{k \geq 0} \frac{e^{-t \lambda} (1+n) \lambda^k}{k!} \left(  (t-n-k-2)^k \theta(t-n-k-2) + (t-n-k-1)^k \theta(t-n-k-1)\nonumber \right ) \\
\fl- \sum_{n \geq 0} \sum_{k \geq 0} \frac{2 e^{-k \lambda -(2+n)\lambda - (t-n-k-2)\lambda  }  \lambda^{k+1}}{(k+1)!} (1+k)(t-n-k-2)^{1+k} \theta(t-n-k-2).
\end{eqnarray}

\begin{figure}
\begin{center}
\includegraphics[width=8cm]{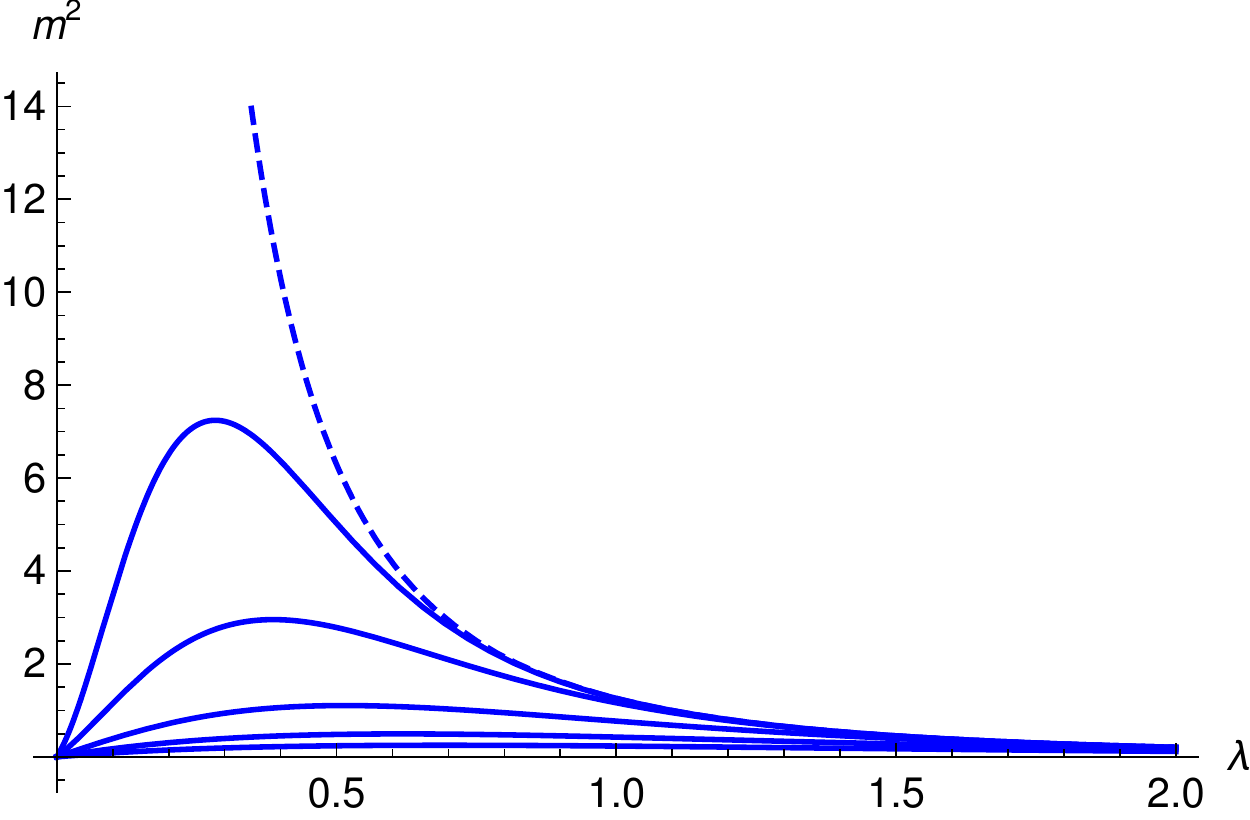}
\end{center}
\caption{Second moment of the number of exiting particles as a function of the intensity for different stopping times, $t_s=20,10,5,3,2$ top to bottom. The dashed line shows the long time
limit, Eq. (\ref{eq:m2u}).}\label{fig:msqt}
\end{figure}

\begin{figure}
\begin{center}
\includegraphics[width=8cm]{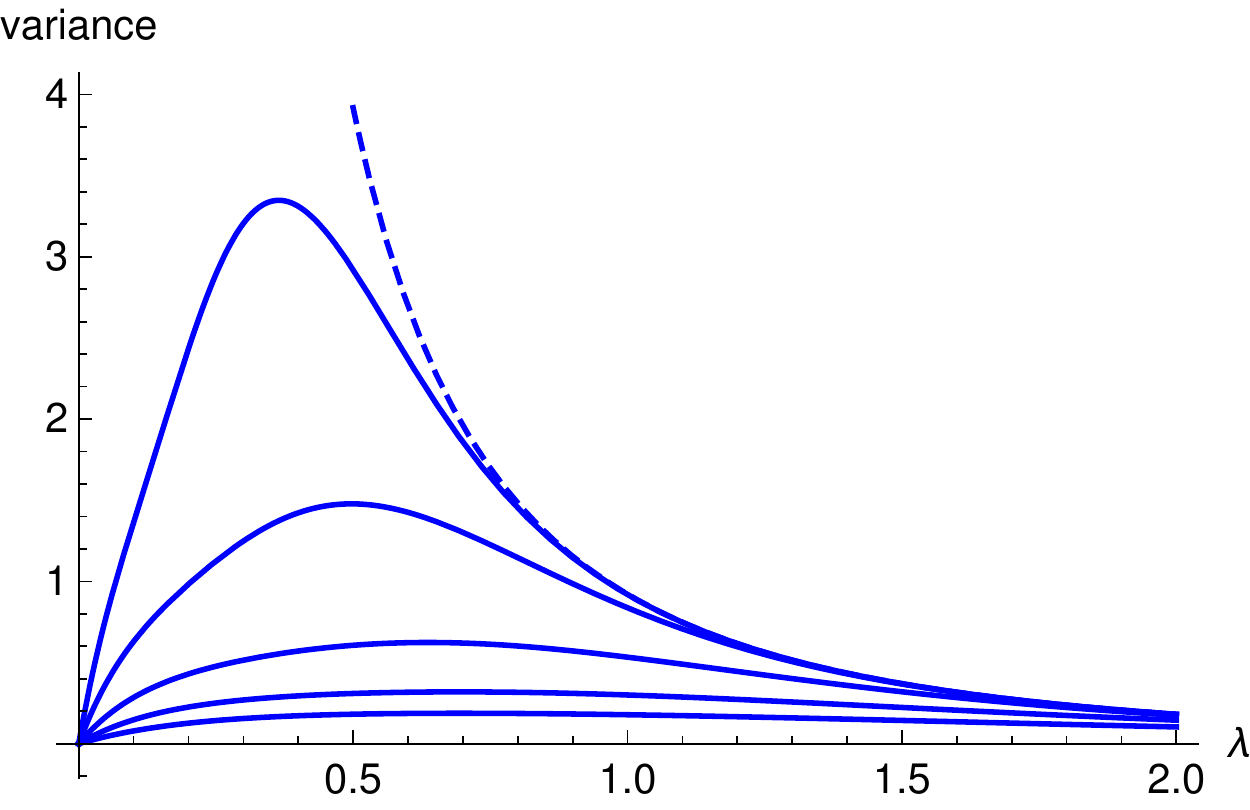}
\end{center}
\caption{Variance of the number of exiting particles as a function of the intensity for different stopping times, $t_s=10,6,4,2,1.5$ top to bottom. The dashed line shows the long time
limit, Eq. (\ref{eq:varlimit}).}\label{fig:variance_intensity}
\end{figure}

\begin{figure}
\begin{center}
\includegraphics[width=8cm]{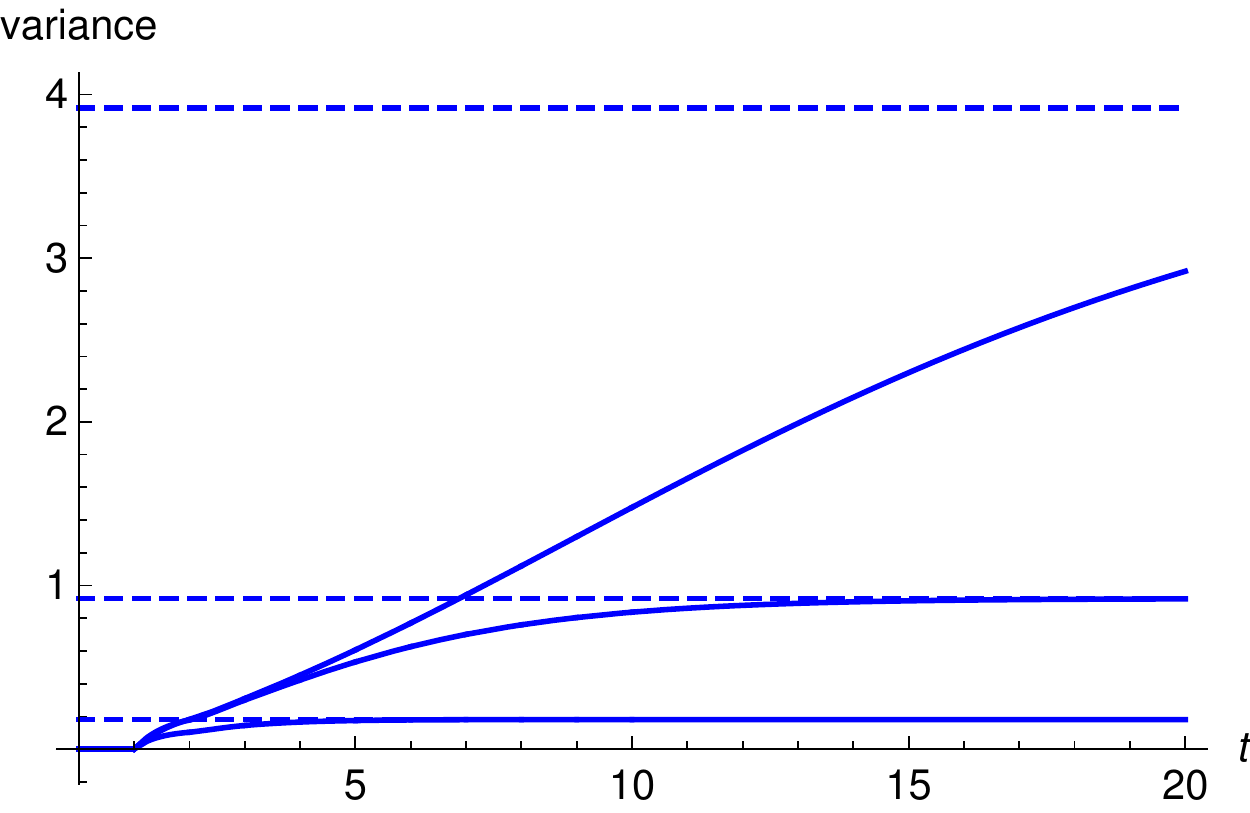}
\end{center}
\caption{Variance of the number of exiting particles as a function of time for different intensities, $\lambda=0.1,0.5,1,2$. The dashed lines show the long time limit, Eq. (\ref{eq:varlimit}).}\label{fig:variance_time}
\end{figure}
\begin{figure}

\begin{center}
\includegraphics[width=8cm]{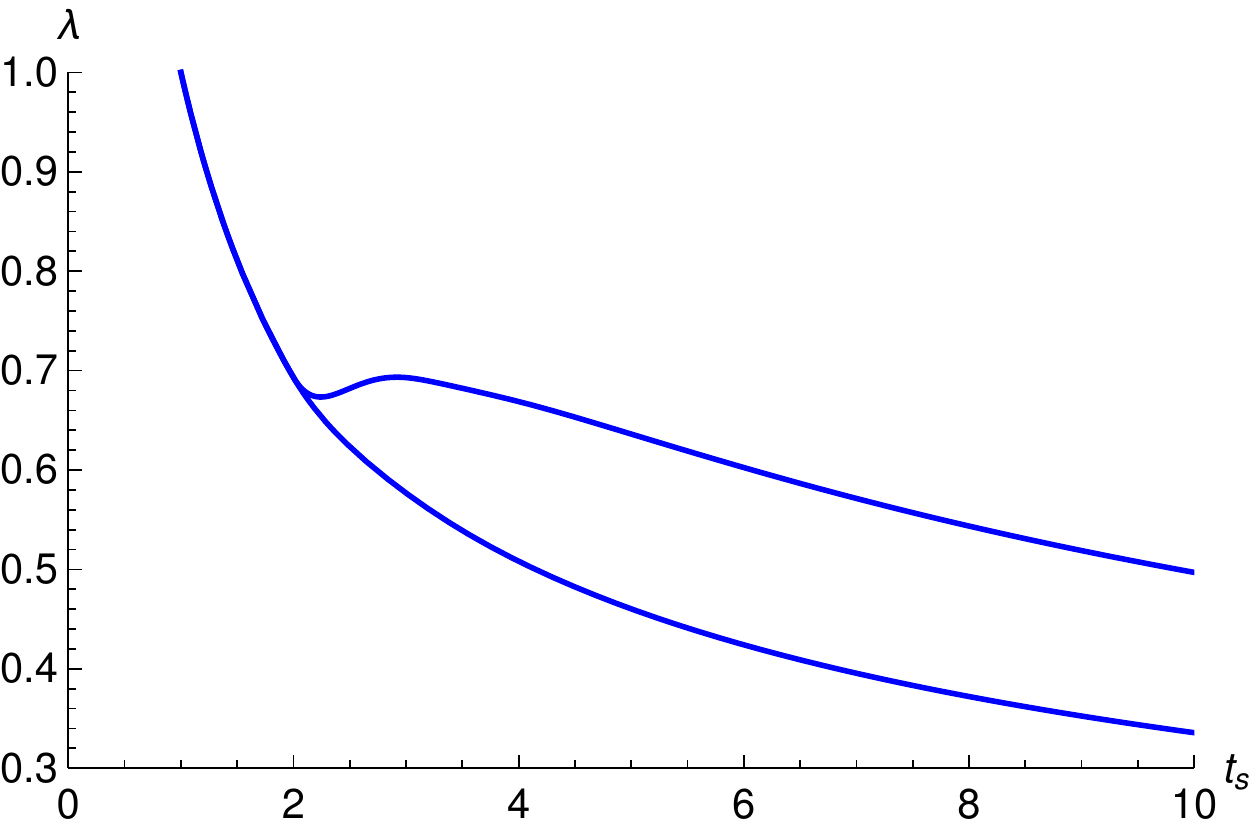}
\end{center}
\caption{Intensity that maximizes the variance (upper curve) and mean of number of exiting particles as a function of the stopping time.}\label{fig:lam_max}
\end{figure}

Numerical results for the second moment and the variance as a function of $\lambda$ 
are shown in Figs.  \ref{fig:msqt} and \ref{fig:variance_intensity}, respectively. 
Like the first moment (Fig. \ref{fig:mout}), the second moment displays a maximum at a finite value of $\lambda$. 
The family of curves at different stopping times approaches the long-time limit, Eq. (\ref{eq:m2u}) as
time increases. The variance also displays a maximum as a function of intensity, see Fig. \ref{fig:variance_intensity}, but the location of the maximum for a given stopping time is displaced to a higher value of the intensity. The time dependent variance for different values of $\lambda$ is shown in Fig. \ref{fig:variance_time}. Finally in Fig. \ref{fig:lam_max} we plot the value of $\lambda$ that maximizes the mean and variance as a function of the stopping time. We observe that
the intensity that maximizes the mean number of exiting particles is a strictly decreasing function of the stopping time. 
The intensity that maximizes the variance of the number of exiting particles, however, has a non-trivial behavior. 
It is the same that maximizes the intensity for $1\le t_s \le 2$, but it then increases to a maximum value for $t_s\approx 3$ and then decreases. For $t_s>2$, the variance is maximized at a higher value of $\lambda$ than the one that maximizes the mean number.  
 
\section{Conclusion}

We have studied the optimization of the throughput of a stream of particles subject to blocking using a circular Markov chain model. 
The sojourn time of a particle contained in an open channel is exponentially distributed with rate $\mu$. If $N$ particles are simultaneously present
the channel is blocked and all newly arriving particles are rejected. After an exponentially distributed blockage time with rate $\mu^*$
all particles forming the blockage simultaneously exit the channel. We presented general expressions for the steady state probabilities and
throughput. For $N=2$ we showed that the steady state throughput assumes a maximum value at finite intensity if $\mu^*/\mu<1/4$. The time
dependent throughput may also display a maximum if $\mu^*/\mu<1/2$. We showed that this behavior is qualitatively different from 
the well-known $M/M/N/N$ queue whose steady state throughput always increases monotonically with the intensity of entering particles. We also 
compared the new model with a previously introduced one with deterministic transit and blockage times. For $N=2$ we found
an exact mapping between the two in the steady state. In future work we plan to apply the circular Markov chain model 
to multi-channel systems where the entering flux is evenly distributed over the open channels \cite{barre2015b}.

In the second part of the article we examined an irreversible blockage model with capacity $N=2$, fixed transit time and an input of
constant intensity that is switched off after a given time, $t_s$. For small stopping times, the mean and variance of the number of exiting particles are
maximized at the same value of the intensity $\lambda$. If $t_s/\tau>2$, the maximum value of the variance occurs at a higher value of the intensity 
than the mean value.

\section{Appendices}

\subsection{Derivation of the Mean First Passage time to the blocked state}

In Cohen \cite{CohenSingle1982} the following formulas can be found for the average 
first hitting times $\nu_{i,j}$ of hitting level $j$, starting from level $i$,
in a birth-death process with birth rates $\lambda_n$ and death rates $\mu_n$ in state $n$:

For $i < j$,
\begin{equation}
\nu_{i,j} = \sum_{n=i}^{j-1} \frac{1}{\lambda_n \pi_n} \sum_{k=0}^n \pi_k.
\label{av_hit_time_i<j}
\end{equation}

For $i = j$ (= average first return time of state $j$),
\begin{equation}
\nu_{i,j} = \frac{\sum_{k=0}^{\infty} \pi_k}{(\lambda_j + \mu_j)\pi_j}.
\label{av_hit_time_i=j}
\end{equation}

For $i > j$,
\begin{equation}
\nu_{i,j} = \sum_{n=j}^{i-1} \frac{1}{\lambda_n \pi_n} \sum_{k=n+1}^{\infty} \pi_k.
\label{av_hit_time_i>j}
\end{equation}

Here, 
\begin{equation}
\pi_0=1, \quad \pi_n = \frac{\lambda_0\lambda_1 \cdots \lambda_{n-1}}{\mu_1\mu_2\cdots \mu_n}.
\label{pi_def}
\end{equation}

In the special case of an $M/M/\infty$ queue we have $\lambda_n=\lambda$ and $\mu_n=n \mu$,
leading to $\pi_n = \frac{\lambda^n}{n! \mu^n} = \frac{\rho^n}{n!}$, where $\rho=\lambda/\mu$.
In this case, we obtain
\begin{equation}
\nu_{0,j} = \sum_{n=0}^{j-1} \frac{1}{\lambda \frac{\rho^n}{n!}} \sum_{k=0}^n \frac{\rho^k}{k!} = \frac{1}{\lambda} \sum_{n=0}^{j-1} n! \sum_{k=0}^n \frac{\rho^{k-n}}{k!}.
\label{m_m_inf_nu_0<j}
\end{equation}
that is Eq. (\ref{eq:mfpt}).

\subsection{Joint probabilities for irreversible blockage at finite time}

In \cite{Talbot2015} we considered the joint probability that
$m$ particles have exited at time $t$ {\it and}  the system is blocked, which we denoted as $h(m,t)$. 
Let us consider the joint probability $g(m,t)$ that $m$ particles have exited the channel
at time $t$ {\it and} that the channel is still open. Clearly

\begin{equation}\label{eq:mdet}
 m(t) = \sum_{k=1}^{\infty}k (g(k,t)+h(k,t))
\end{equation}
and  $f(k,t)$ introduced in Sec. \ref{sec:irrevmodel} is simply $f(k,t)=g(k,t)+h(k,t)$.

The time evolution of $g(n,t)$ is given by
\begin{eqnarray}
\fl \frac{dg(0,t)}{dt}&=&-\lambda \int_0^{{\rm min}(t,1)}dt_1 \lambda e^{-\lambda t_1} q_s(0,t-t_1) -\lambda e^{-\lambda } q_s(0,t-1)\nonumber\\
\fl  \frac{dg(1,t)}{dt}&=&\lambda e^{-\lambda }q_s(0,t-1)-\lambda \int_0^{{\rm min}(t-1,1)}dt_1 \lambda e^{-\lambda(t_1+1)} q_s(1,t-1-t_1)-\lambda e^{-2\lambda }q_s(1,t-2)\nonumber\\
\fl \frac{dg(n,t)}{dt}&=&\lambda e^{-2\lambda }q_s(n-1,t-2)-\lambda \int_0^{{\rm min}(t-1,1)}dt_1 \lambda e^{-\lambda(t_1+1)} q_s(n,t-1-t_1)
 -\lambda e^{-2\lambda }q_s(n,t-2).\nonumber\\
\end{eqnarray}
The time derivative of $g(n,t)$ is given as the sum of a gain term and two loss terms. The gain term is the probability 
density that the $nth$ particle exits at time $t$ and that the channel is still open. This corresponds to the event where  the $nth$ particle enters at $t-1$ and that $n-1$ particles have 
already exited the channel.  The first loss term corresponds to a particle which blocks the channel at time $t$ knowing that $n$ particles already exited.
This means that a particle is still in the channel at $t$ and a new one entering at time $t$ blocks the channel. 
The last term corresponds to the exit of the $nth$ particle at time $t$ with a channel still open.
The boundary term for $n=0$ does not require a time lag  in the two  loss terms because for $t>1$ a particle can enter without clogging 
the channel. A similar argument applies for $n=1$ to the gain term.
Defining the Laplace transform as
\begin{equation}
 \tilde{g}(u)=\int_0^\infty dt e^{-ut} g(t),
\end{equation}
the differential equations become
\begin{equation}
u\tilde{g}(0,u)-1 =-\lambda\frac{\lambda+ue^{-(\lambda+u)}}{\lambda+u}  \tilde{q}_s(0,u).
\end{equation}
Knowing that $\tilde{q}_s(0,u)=\frac{1}{\lambda +u}$, the inverse Laplace transform of $\tilde{g}(0,u)$ is 
\begin{equation}
 \tilde{g}(0,u)=\frac{1}{u}-\frac{\lambda}{u}\left(\frac{\lambda+ue^{-(\lambda+u)}}{(\lambda+u)^2}\right),
\end{equation}
which gives
\begin{equation}
 g(0,t)=(1+\lambda t -\theta(t-1)\lambda(t-1))e^{-\lambda t}.
\end{equation}

To go further, let us recall that the generating function for  $h(k,t)$, $G_h(z,t)=\sum_k z^kh(k,t)$, is given in Laplace space by (see \cite{Talbot2015})
\begin{equation}
 \tilde{G}_h(z,u)=\frac{\lambda^2(1-e^{-(\lambda+u)})}{u(\lambda+u)(u+\lambda(1-ze^{-(\lambda+u)})}.
\end{equation}
Introducing a generating function for $g(k,t), G_g(z,t)=\sum_k z^kg(k,t)$, one can express the number of exiting particles as
\begin{equation}\label{eq:mdetgen}
 m(t)=\frac{\partial G_g}{\partial z}(1,t)+\frac{\partial G_h}{\partial z}(1,t).
\end{equation}
By combining the differential equations of $g(k,t)$ and $h(k,t)$, one obtains
\begin{equation}\label{eq:generatingall}
 \frac{\partial (G_g(z,t)+G_h(z,t))}{\partial t}=(z-1)\lambda e^{-\lambda}q_s(0,t-1)+(z-1)\lambda e^{-2\lambda}(G(z,t-2)-q_s(0,t-2)),
\end{equation}
where $G(z,t)$ is the generating function of $q_s(k,t)$.
By taking the Laplace transform of Eq.(\ref{eq:generatingall}), one has
\begin{equation}\label{eq:laplacegene}
 u(\tilde{G}_g(z,u)+\tilde{G}_h(z,u))-1=(z-1)\frac{\lambda e^{-(\lambda+u)}}{\lambda+u} +(z-1)\lambda e^{-2(\lambda+u)}(\tilde{G}(z,u)-\frac{1}{\lambda+u}).
\end{equation}
Taking the partial derivative of Eq.(\ref{eq:laplacegene}) with respect to $z$ and using Eq.(\ref{eq:mdetgen}), one obtains
\begin{equation}\label{eq:mdeu3}
 \tilde{m}(u)=\frac{1}{u}\left(\frac{\lambda}{\lambda+u}e^{-(\lambda+u)}(1-e^{-(\lambda+u)})+\lambda e^{-2(\lambda+u)}\tilde{G}(1,u)\right).
\end{equation}
Let us recall that (see Ref.\cite{Talbot2015})
\begin{equation}\label{eq:generatingG}
 \tilde{G}(1,u)=\frac{1}{\lambda+u}\left[1+\frac{\lambda}{\lambda+u-\lambda e^{-(\lambda+u)}}\right].
\end{equation}
Finally, inserting Eq. (\ref{eq:generatingG}) in Eq. (\ref{eq:mdeu3}), the Laplace transform of $m(t)$ is given by 
\begin{equation}\label{eq:mdeu2}
  \tilde{m}(u)=\frac{\lambda}{u} \frac{e^{-(\lambda+u)}}{\lambda+u-\lambda e^{-(\lambda+u)}}.
\end{equation}

Note that Eq. (\ref{eq:mdeu2}) can be written  as
\begin{equation}
  \tilde{m}(u)=\frac{\tilde{j}(u)}{u}
\end{equation}
and one now recovers that $m(t)=\int_1^t dt' j(t') $.
One also checks that $m(\infty)=\lim_{u\rightarrow 0}u\tilde{m}(u)=\frac{e^{-\lambda}}{1-e^{-\lambda}}$ as expected.


\section*{References}

\providecommand{\newblock}{}

\end{document}